\documentclass[fleqn,10pt]{wlscirep}

\usepackage{graphicx}% Include figure files
\usepackage{dcolumn}% Align table columns on decimal point
\usepackage{bm}% bold math

\usepackage{braket}
\usepackage{physics}
\usepackage{color}

\begin{document}
 %\preprint{APS/123-QED}

  \title{Giant excitonic absorption and emission in two-dimensional group-III nitrides}% Force line breaks with \\
\author[1]{Maria Stella Prete}
\author[1]{Davide Grassano}
\author[1,*]{Olivia Pulci}
\author[2]{Ihor Kupchak}
\author[3]{Valerio Olevano}
\author[4]{Friedhelm Bechstedt}
\affil[1]{ Dipartimento di Fisica, Universit\`{a} di Roma Tor Vergata, and INFN, Via della Ricerca Scientifica 1, I-00133 Rome, Italy}
\affil[2]{
V.E. Lashkaryov Institute of Semiconductor Physics, National Academy of Sciences of Ukraine, Kyiv, Ukraine}
\affil[3]{CNRS, Institut Neel, 38042 Grenoble, France}
\affil[4]{ IFTO, Friedrich Schiller Universit\"at, Max-Wien Platz 1, D-07743 Jena, Germany}
\affil[*]{olivia.pulci@roma2.infn.it}

  \date{\today}% It is always \today, today,

  \begin{abstract}
    Absorption and emission of pristine-like semiconducting monolayers of BN, AlN, GaN, and InN are here systematically studied by \emph{ab-initio} methods.
    We calculate the absorption spectra for in-plane and out-of-plane light polarization including quasiparticle and excitonic effects.
    Chemical trends with the cation of the absorption edge and the exciton binding are  discussed in terms of the band structures.
    Exciton binding energies and localization radii are explained within the Keldysh model for excitons in two dimensions.
    The strong excitonic effects are due to the interplay of low dimensionality, confinement effects, and reduced screening.
    We find exciton radiative lifetimes ranging from tenths of picoseconds (BN) to tenths of nanoseconds (InN) at room temperature, thus making 2D nitrides, especially InN,  promising materials for light-emitting diodes and high-performance solar cells.
  \end{abstract}

  \maketitle

  %\tableofcontents
 \section*{\label{sec1}Introduction}
  % section{\label{sec1}Introduction}
    Since the discovery of graphene, the research on alternative two-dimensional (2D) materials has gained enormous interest.
    Silicene, germanene, phosphorene, and transition metal dichalcogenides are just a few examples of classes of novel low-dimensional systems which may serve as building blocks for efficient nanoelectronic and nanooptical devices \cite{Wang.Kalantar-Zadeh.ea:2012:NN,Balendhran.Walia.ea:2015:S,Xia.Wang.ea:2014:NP,grazianetti2018optical}.
    On the other hand, bulk group-III nitrides such as GaN, AlN and InN are most important materials for solid state lighting, as witnessed by the Nobel prize awarded in 2014 to  Akasaki, Amano and Nakamura \cite{Nakamura:2015:RMP}.
    The possibility to play with dimensionality to enhance the already exceptional properties of bulk nitrides has caused an avalanche of attempts to grow  2D nitrides, although the strong tendency for $sp^3$ bonding makes their preparation extremely difficult.
    Only monolayer BN can, in principle, be easily prepared by exfoliation from hexagonal bulk BN \cite{nagashima1995electronic}.
    Promising experimental attempts to obtain 2D AlN and GaN have recently appeared \cite{tsipas2013evidence,alame2016preparation,wang2019experimental}.
    Measurements of GaN sheets encapsulated in graphene seem to suggest a fundamental optical gap of about 5 eV \cite{al2016two}.
    Gaps even larger than the  BN  one have been found for low-dimensional AlN structures \cite{wang2019experimental}.
    These successful experiments pave the way to new photovoltaic and optical nanodevices based on 2D nitrides \cite{Kecik.Onen.ea:2018:APR}.

    Experimental activities to prepare monolayer and few-layer group-III nitride sheets are accompanied by theoretical studies.
    For graphene-like, hexagonal 2D AlN and GaN, their dynamical stability has been explored and verified by phonon or even molecular-dynamics calculations \cite{zhuang2013computational,Sahin.Cahangirov.ea:2009:PRB,Onen.Kecik.ea:2016:PRB,peng2018room,Shu.Niu.ea:2019:ASS}.
    The progress toward theoretical growth predictions has been recently summarized in a review paper \cite{Kecik.Onen.ea:2018:APR}.
    Although the crystal structure of freestanding AlN and GaN is under debate, most studies favor a planar, honeycomb structure \cite{zhuang2013computational,Kecik.Onen.ea:2018:APR,bacaksiz2015hexagonal,guilhon2019out}.

    The recent progress in achieving growth of ultrathin 2D AlN and GaN layers on  substrates, as well as the exfoliation of BN monolayers, suggest the possibility to  prepare also 2D InN.
    Therefore, a theoretical understanding of the monolayer properties of BN, AlN, GaN, and InN, in particular of their electronic structure as well as optical absorption and emission properties, may push forward the experimental activities on 2D nitrides and their possible optoelectronic applications.
    Several studies have been focused on the tunable electronic gap \cite{Sahin.Cahangirov.ea:2009:PRB,Onen.Kecik.ea:2016:PRB,prete2017tunable,Wang.Wu.ea:2018:JPCC,peng2018room,Shu.Niu.ea:2019:ASS}.
    Excitonic properties have been recently investigated for GaN \cite{Shu.Niu.ea:2019:ASS,peng2018room,Shu.Niu.ea:2019:ASS,sanders2017electronic} and InN \cite{Liang.Quhe.ea:2017:RA,Prete.Pulci.ea:2018:PRB}, in addition to the activities on the well-known BN layer (\cite{wirtz2006excitons,guilhon2019out} and references therein).

    In this letter, we tackle this goal through a systematic \emph{ab-initio} study of the electronic and accompanying optical properties of 2D group-III nitrides BN, AlN, GaN, and InN.
We focus on the absorption and emission spectra near the onsets for in- and out-of-plane light polarization.
    The dominating bound excitons are studied in detail, in particular the influence of spatial confinement effects and reduced screening of the electron-hole interaction.
    For the exciton ground state we calculate the binding energy and the excitonic radius and compare the results with those of model calculations.
    The emission is then characterized by the determination of the excitonic radiative lifetime.

  \section*{\label{sec3}Results}
    \subsection*{\label{sec3a}Geometry and electronic properties}
      2D sheets of group-III nitrides have as equilibrium geometry a flat honeycomb structure similar to graphene \cite{cahangirov2009two} but with reduced D$_{3h}$ symmetry.
      As in graphene and BN, also in AlN, GaN and InN the first-row element N tends toward a $sp^2$ hybridization and in-plane III-N bonds.
      Therefore, the geometries are only characterized by the lattice constant $a$.
      The lattice parameters, listed in Table~\ref{tab:geo}, follow a clear chemical trend with increasing atomic number of the cation.
      The lattice constants are smaller than those for the 3D bulk counterparts, consistently with the $sp^2$ bonding being stronger than the $sp^3$ one.

      The electronic properties of the 2D III-N monolayers, conversely to graphene, do not show any  Dirac cone  in the electronic band structure \cite{prete2017tunable,Wang.Wu.ea:2018:JPCC,Shu.Niu.ea:2019:ASS,cahangirov2009two}.
      The band structures, see Supplemental Material (SM) Fig.~SM1, are characterized by indirect gaps with the valence band maximum (VBM) at $K$ and the conduction band minimum (CBM) at $\Gamma$.
      Apart from BN with a lowest direct band gap at $K$, the other nitrides exhibit a direct gap at $\Gamma$.
      While BN and AlN tend to be indirect semiconductors, GaN and InN are direct 2D crystals \cite{prete2017tunable}.
      Table~\ref{tab:geo} indicates that the valence band maxima at $\Gamma$ and $K$ are not far away from each other.
      However, the direct and indirect gaps of 4.53 and 4.57 eV are close to each other.
      Indirect gaps of 4.55 eV \cite{Onen.Kecik.ea:2016:PRB}, 4.38 \cite{peng2018room} and 4.44 eV \cite{Shu.Niu.ea:2019:ASS} have been computed.
      Fully self-consistent QP calculations \cite{peng2018room} seem to strengthen the tendency for a direct gap at $\Gamma$ but with a larger value of 5.39 eV.
      Experimental values of GaN layers embedded in AlN cover a wide range of 4.76 - 5.44 eV \cite{bayerl2016deep} not too far from all the QP gaps.
      Moreover, the values in Table~\ref{tab:geo} are close to those found in the literature for BN \cite{guilhon2019out}, AlN and InN \cite{prete2017tunable}.

      The gap values $E_g$, listed in Table~\ref{tab:geo} are due to VBM composed primarily of N2$p_z$ orbitals, while the CBM is a hybridization of group-III s and some N2s states.
      The gaps arise from the presence of two different kind of atoms, bearing  different electronegativity, in the two hexagonal sublattices. 
      A charge transfer occurs from the cation (B, Al, Ga, In) to the anion (N) with a subsequent opening of the electronic QP gaps, which span from about 7 eV for 2D BN to 1.7 eV for 2D InN, respecting the chemical trend.
      The electronic gaps shown by 2D III-N monolayers are much larger than those of their 3D bulk counterparts (see collections in \cite{wirtz2006excitons,Carvalho.Schleife.ea:2011:PRB,Sahin.Cahangirov.ea:2009:PRB}).
      Quantum confinement of the electrons is mainly responsible for this behavior.
      In addition, the reduced 2D screening also modifies the QP corrections.

 \subsection*{\label{sec3b}Excitonic effects in optical spectra}

\begin{figure}[ht]
        \includegraphics[scale=0.8]{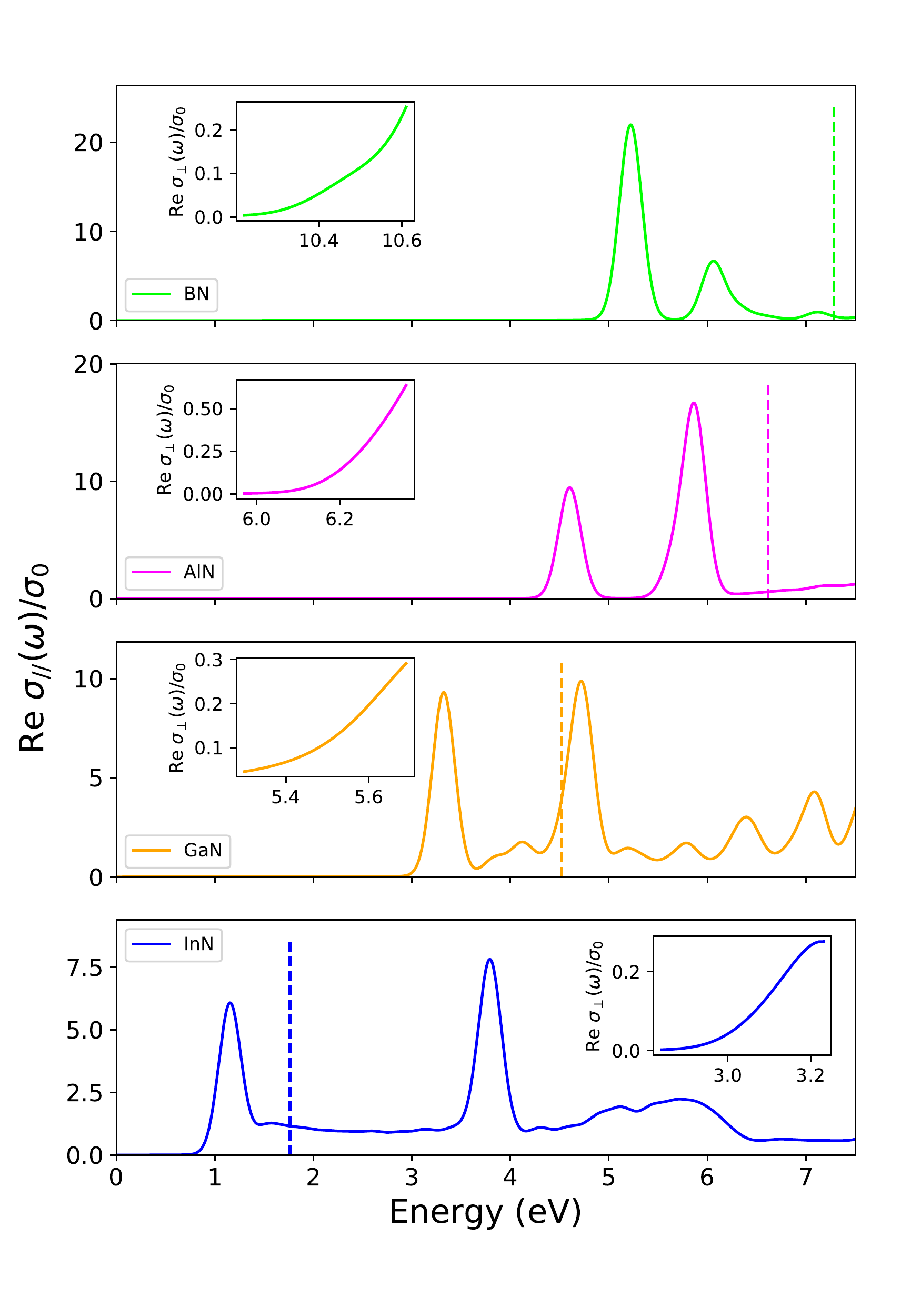}
					\caption{
						Real part of the dynamical optical conductivity ${\text Re} \sigma_{\|}(\omega)$ for in-plane polarization of (a) BN, (b) AlN, (c) GaN, (d) InN.
            The insets show the onset of absorption for out-of-plane polarization.
						The spectra are normalized to the dc conductivity $\sigma_0$.
						The direct QP gap is indicated by a vertical dashed line.
						}\label{fig:sheet_polarizability}
      \end{figure}

      The absorption spectra including the QP and excitonic effects are displayed in Fig.~\ref{fig:sheet_polarizability}.
      Indeed the real part of the optical conductivity represents the absorbance for vanishing reflectance.
      The inclusion of electron-hole interaction through BSE leads to a redshift of absorption spectra and seems to compensate quasiparticle effects, restoring a qualitative agreement with the position of DFT spectra but not their lineshape (see Figs.~SM2 to SM5).
      For in-plane light polarization
      strong peaks appear due to bound excitons, well below the fundamental direct gaps.
      For GaN and InN, and especially InN, such bound exciton peaks also occur above the fundamental QP gap belonging to high-energy interband transitions and, hence, being resonant with the continuum of scattering states of lower valence-conduction band pairs.
      The striking difference between the two light polarizations is the significant blueshift of the absorption spectra ${\text Re} \sigma_\bot(\omega)$ mainly due to local-field or depolarization effects described by the electron-hole exchange interaction in the two-particle Hamiltonian $H_{\rm exc}$ \cite{matthes2016influence,moody2016exciton,guilhon2019out,marinopoulos2004optical}.
      Also their intensities are reduced, in particular for BN.

      In Fig.~\ref{fig:sheet_polarizability} we focus on the onset of the absorption spectra for in-plane light polarization.
      For BN the onset is characterized by three low-energy peaks of decreasing intensity.
      For the first exciton at 5.2 eV, usually identified as optical gap $E^{\rm opt}_g$ reduced by the binding energy $E_b$ with respect to the QP gap $E_g$, we find a large exciton binding energy of 2.0 eV.
      A second and a weak third peak appear at 6.1 eV and 7.2 eV (see Fig.~SM2), respectively.
      The first structure originates from transitions at the fundamental gap at $K$ between the highest valence band and the lowest conduction band.
      The second exciton peak has a similar origin (see Fig.~SM1).
      The first two band-edge excitons are related to $\pi\rightarrow\pi^*$ transitions.
      The third peak belongs mainly to transitions near $\Gamma$ with some hybridization with $\sigma$ and $\sigma^*$ states.
      Although the position of the peaks is slightly different, our value of the binding energy is in agreement with other predictions \cite{wirtz2006excitons,guilhon2019out,Berseneva.Gulans.ea:2013:PRB}.
      The AlN in-plane conductivity exhibits  two prominent structures of increasing intensity, located at 4.6 eV and 5.9 eV.
      The binding energy of the first exciton is  1.9 eV.
      In a previous theoretical work \cite{bacaksiz2017h} three excitons peaks at 4.12 eV, 4.97 eV and 5.4 eV were found, with a similar binding energy of 1.88 eV.
      Also for GaN we register the presence of two main peaks for the in-plane component of the BSE sheet polarizability near the absorption edge.
      The first peak appears at 3.3 eV while the second at 4.7 eV (see also SM), similarly to other calculations \cite{Shu.Niu.ea:2019:ASS}.
      The first exciton presents a large binding energy of 1.2 eV in agreement with Refs. \cite{Shu.Niu.ea:2019:ASS,peng2018room,sanders2017electronic}.
      According to Fig.~SM1(c),the first exciton is mainly built by allowed, mainly $\sigma \rightarrow \sigma ^\star$, optical transitions between the highest two valence bands and the lowest conduction band near $\Gamma$.
      The second broader excitonic feature in Fig.~\ref{fig:sheet_polarizability} and SM4(a) with partial resonant character, is mainly related to the lowest energy $\pi \rightarrow \pi ^ \star$ transitions near $K$, combined with transitions along the $K \rightarrow M$ line in the BZ, because of the flatness of the lowest conduction and highest valence bands.
      However, also contributions between the highest valence bands and the second lowest conduction band near $\Gamma$, with $\sigma \rightarrow \pi ^\star$ character, do contribute.
      Two bound excitons appear for InN in the BSE optical conductivity for in-plane light polarization below photon energies of 6~eV.
      The first exciton, at 1.2 eV, has a binding energy of 0.6 eV and originates VBM-CBM transitions near $\Gamma$.
      A second bound but resonant exciton pea,k visible at 3.9 eV, arises from VBM-CBM transitions near $M$ and $K$ at the BZ boundary.

    \subsection*{Exciton binding and localization}\label{sec:3c}
\begin{figure}[ht]
	      \includegraphics[scale=0.4]{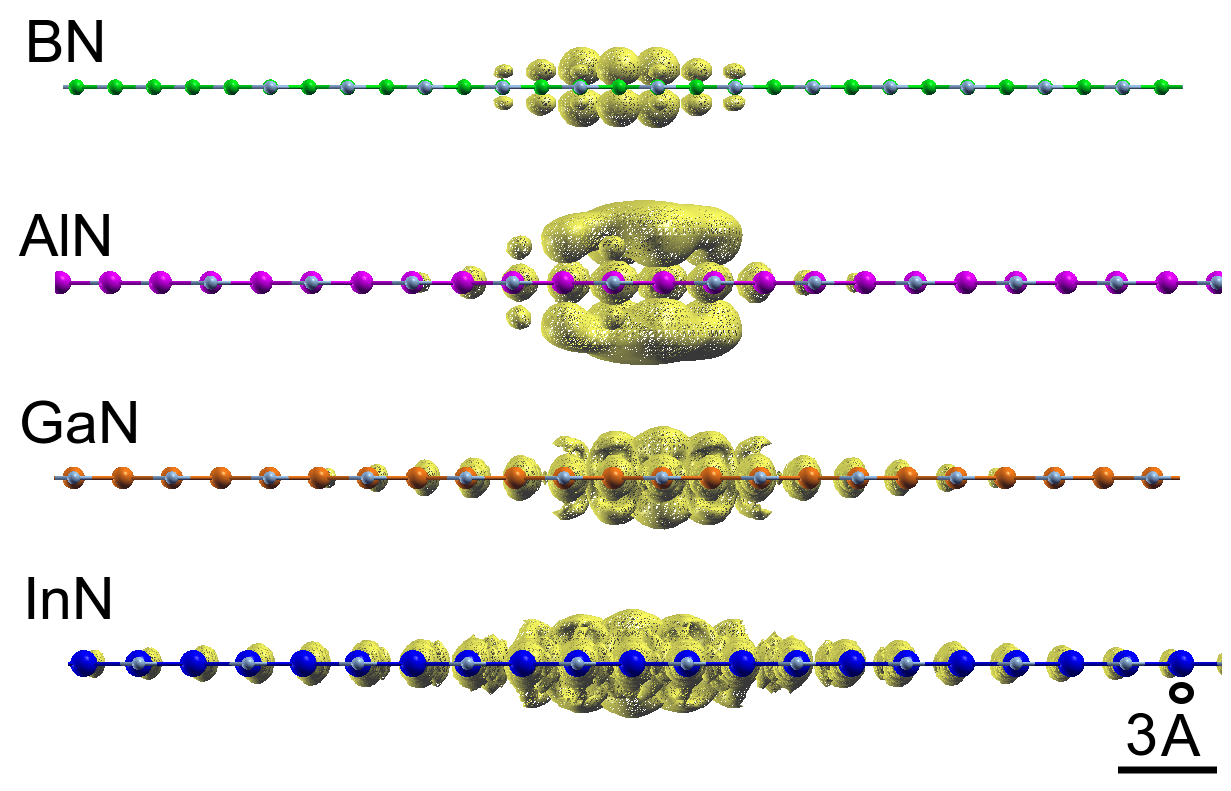}
	      \caption{
		      Exciton wavefunctions for the 2D nitrides.
		      In all cases the hole has been fixed near a nitrogen atom (light blue).
		      }\label{fig:exc_wf}
      \end{figure}

\begin{figure}[ht]
	      \includegraphics[scale=0.8]{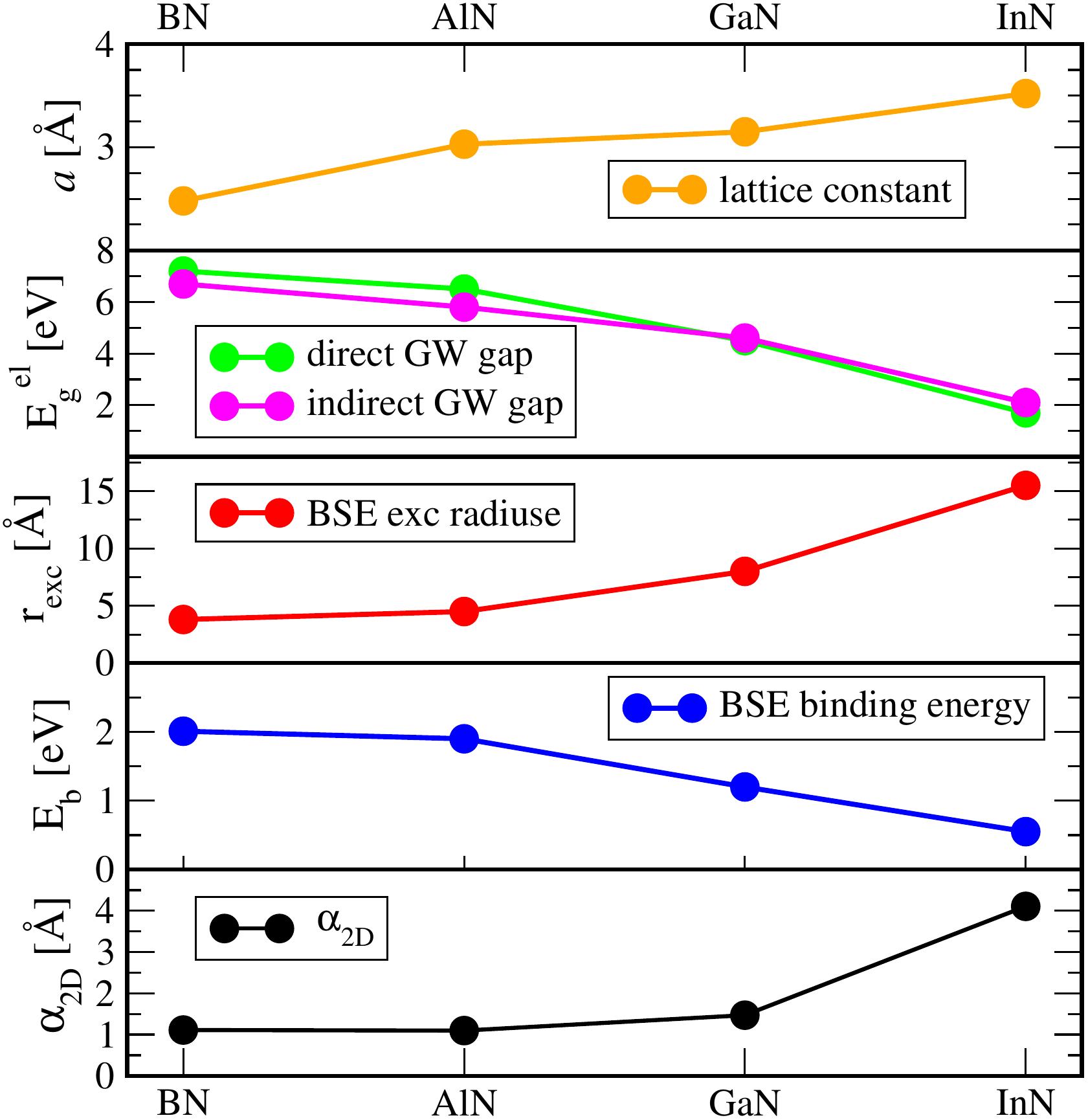}
	      \caption{
		      \emph{Ab-initio} results for the lattice constants, direct and indirect QP gaps, exciton radius, binding energy of the excitons and the 2D static polarizability.
		      }\label{fig:trend}
      \end{figure}

      In all 2D nitrides the first strong excitonic peak originates from transitions at the fundamental direct gap between the highest valence band and the lowest conduction band near $\Gamma$ ($K$ in the BN case).
      Their binding energies $E_b$ are strongly related to the exciton radii $r_{exc}$ as expressed by the chemical trends in Table~\ref{tab:geo} and Fig.~\ref{fig:trend}.
      Thereby, the exciton radius has been derived {\it ab-initio} from the expectation value with the computed exciton wavefunctions.
       The corresponding exciton wave functions are plotted in Fig.~\ref{fig:exc_wf}. % Fig.~\ref{fig:exc_wf}.
  They represent the probability to find the electron, having fixed the position of the hole near one nitrogen atom.
      Their localization around the chosen hole position clearly indicates the 2D character of the excitons, which extends above and below the basal plane just for a length of the order of the lattice parameter.
      The lateral electron-hole distance covers instead a wider range in Fig.~\ref{fig:exc_wf}.
      By defining the excitonic radius $r_{exc}$ as the first moment
      \begin{equation}
    	  r_{exc}=\int d^3 {\bf r}_e|{\bf r}_e|\psi^*({\bf r}_h,{\bf r}_e)\psi({\bf r}_h,{\bf r}_e),
      \end{equation}
      where $\psi({\bf r}_h,{\bf r}_e)$ denotes the exciton eigenstate $|S{\bf Q}\rangle$ for the state $S=0$ and ${\bf Q}=0$ in real-space representation,
      we find an excitonic radius as small as 3.8 \AA~ for 2D BN, and as large as 15.5 \AA~ for 2D InN.
      Comparing the values $r_{exc}$  with the lattice constants $a$ in Table~\ref{tab:geo}, a clear tendency for the formation of 2D excitons in the ground state from BN to InN becomes visible.
      The averaged electron-hole distances $r_{\rm exc}$ in BN and AlN are not significantly larger than the lattice parameters.
      Their lowest energy excitons may be therefore interpreted to possess a Frenkel-like character.
      Electron and hole in the excitons are found close to nearest-neighbor atoms in the plane with the highest probability.
      In the case of GaN and InN, however, the larger electron-hole distances $r_{\rm exc}$ suggest a character closer to that of 2D hydrogen-like Wannier-Mott excitons.
      In any case, the excitonic features in the 2D nitrides are much stronger compared to their 3D counterparts \cite{Carvalho.Schleife.ea:2011:PRB}, because of two main reasons: the confinement of the motion of electrons and holes in the 2D honeycomb structure and the weak 2D screening of the electron-hole attraction.

      The chemical trends of the exciton parameters $E_b$ and $r_{\rm exc}$  are qualitatively in line with the fundamental QP gaps $E_g$ and the lattice constant $a$ (see Table~\ref{tab:geo}).
      As summarized in Fig.~\ref{fig:trend}, with increasing cation atomic number we find an increasing lattice constant, accompanied by a decrease of the QP gaps.
      The exciton binding energy $E_b$ qualitatively follows the gap energy, which characterizes the intrinsic screening in the sheet.
      Using the polarizability and mass parameters $\alpha _{2D}$, $m_e$ and $m_h$ from Table~\ref{tab:geo}, the variational approach to 2D excitons \cite{Keldysh:1979:JETPL,pulci2015excitons} yields model values for $E_b$ and $r_{exc}$ which are also listed in the same table.
      They clearly demonstrate that neither the limit of the unscreened 2D Coulomb potential, nor the logarithmic behavior for the electron-hole attraction \cite{Keldysh:1979:JETPL}, which represent the two opposite limits of the model Hamiltonian, are fulfilled for 2D nitrides.
      Rather,
      the electron-hole distance $r_{\rm exc}$ nearly scales with the lattice constant in Table~\ref{tab:geo}.
      A quantitative relationship between the two exciton parameters can be approximately given by $E_b\sim\hslash^2/(2\mu r^2_{\rm exc})$ with $\mu$ as the reduced  exciton mass.
      The prefactor varies between 2.9 and 4.4, thereby indicating that the true \emph{ab-initio} computed excitons are not too far from a 2D hydrogen-like model, which would suggest a prefactor 4.
   % \subsection*{\label{sec3c}Model 2D excitons}
      In other words, the  intermediate range is realized, i.e., neither the Wannier-Mott nor the logarithmic behavior gives a reasonable description of the screened potential.
      The reliability of the analytic method, previously tested on graphane, silicane, and germanane in \cite{pulci2015excitons}, is demonstrated  by comparing  the \emph{ab-initio} and 2D model results in Table~\ref{tab:geo}.
      For both exciton binding energy and exciton radius the chemical trends and the absolute values are in astonishing agreement.
      The energy deviations are only of the order of $0.1-0.2$~eV and the radii differences are of the order or less of 20\%.
      For the strongest bound exciton in BN practically the same values are obtained.
      The rough description  of the lowest exciton as a 2D 1$s$ variational state seems hence to cover the correct physics.

\begin{figure}[ht]
	      \includegraphics[scale=0.8]{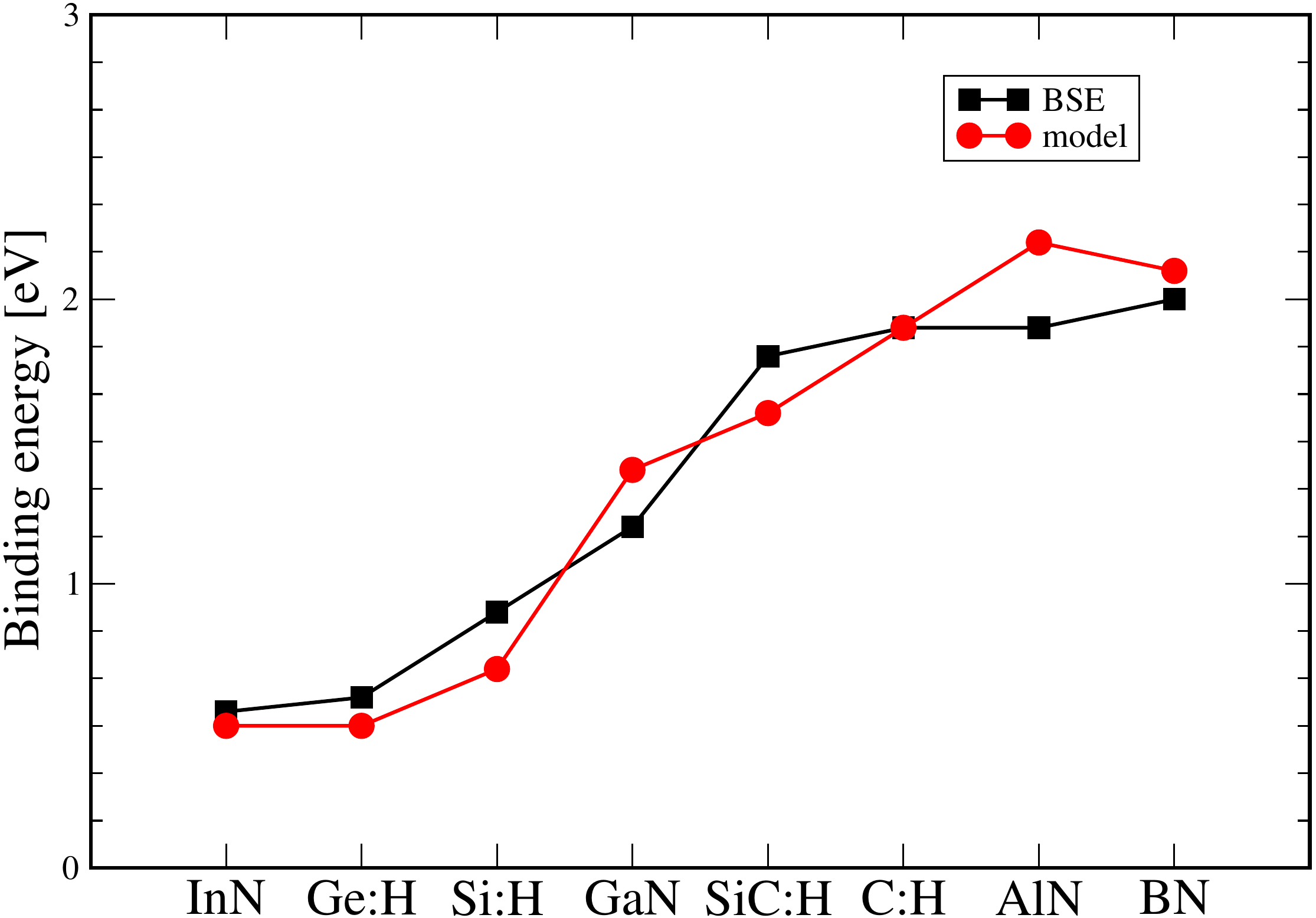}
	      \caption{
  	      Comparison of the exciton binding energies calculated \emph{ab-initio} by solving the BSE (black squares), and by solving the 2D excitonic model (red circles).
  	      }\label{fig:binding_model}
      \end{figure}

      Analyzing the variational results in Table \ref{tab:geo} we can affirm that the model solutions are in good agreement with the results of full \emph{ab-initio} but cumbersome BSE calculations.
      The general validity of this conclusion is underlined in Fig.~\ref{fig:binding_model} by the results for the hydrogenated group-IV materials, graphane, silicongraphane, silicane, and germanane \cite{pulci2015excitons,Pulci.Gori.ea:2012:EPL,Gori.Pulci.ea:2012:APL}, which are in line with the findings for nitride monolayers.

    \subsection*{\label{sec3d}Exciton radiative lifetimes}
      Important material characteristics for light-emitting diodes and laser devices are the rates of different electron-hole recombination processes.
      Not much is known at the moment about these processes in 2D nitrides.
      Here, we focus on the exciton radiative lifetimes and compare the results with the corresponding lifetimes of 3D nitrides.
      Again, we consider the lowest-energy excitons $S=0$ and ${\bf Q}=0$ to find the radiative decay time \eqref{eq:tau$_s$(0)} or its thermal average \eqref{eq:tau$_s$}.
      The lifetimes $\tau_S(0)$ of BN, AlN, and GaN are very similar to each other, while for 2D InN  it is significantly larger (see Table~\ref{tab:geo}).
      The lifetimes and the exciton binding energies in Table~\ref{tab:geo} show opposite trends, in accordance with the Heisenberg uncertainty principle.
      %\begin{table}[ht]
    %    \begin{tabular}{|c|c|c|c|}
     %     \hline \hline
     %          & $\tau_S(0)$ (fs) &$ \langle \tau_S \rangle$(ps) \\ \hline
     %     BN   &  29 (13)         & 55 (14)                      \\ \hline
     %     AlN  &  33 (1)          & 67 (18)                      \\ \hline
     %     GaN  &  35 (18)         & 73 (36)                      \\ \hline
     %     InN  &  55 (58)         & 422 (445)                    \\ \hline \hline
     %   \end{tabular}
     %   \caption{
     %     Calculated exciton radiative lifetimes for 2D III-nitrides $\tau_S(0)$ (\ref{eq:tau$_s$(0)}) and $\langle \tau_S\rangle$ (\ref{eq:tau$_s$}) for room temperature and bound excitons in the ground state $S=0$ and ${\bf Q}$=0.
     %     The values in parenthesis have been estimated within the effective mass approximation but \emph{ab-initio} calculated matrix elements.
     %     }\label{tab:lifetime}
     % \end{table}

      The recombination times of the lowest energy excitons, are by more than one order of magnitude smaller than the corresponding lifetimes of excitons of transition metal dichalcogenides at the corner points $K$ of the hexagonal BZ \cite{palummo2015exciton}.
      The main reason for this difference is the stronger transition matrix elements in the case of the nitrides.
      The radiative lifetimes of the lowest-energy excitons clearly indicate the excellent emission properties of 2D nitrides.
      They seem to be more appropriate for LED and laser applications than the transition metal dichalcogenides.

      The averaged lifetimes for excitons in the lowest-energy exciton band $\langle \tau_S\rangle$ at room temperature still give rise to shorter values than the transition metal dichalcogenides \cite{palummo2015exciton}.
      In general, they possess an averaged radiative lifetime $\langle\tau_S\rangle$, which is three orders of magnitude larger than $\tau_S(0)$.
      This is a consequence of the much faster recombination of the excitons with vanishing translational energy.
      Comparing the averaged lifetimes $\langle \tau_S\rangle$ at room temperature in Table \ref{tab:geo} with the values for 3D nitrides \cite{dmitriev1999rate,jhalani2019first},  differences of one to two orders of magnitude are observed.
      This fact, again, underlines the outstanding emission properties of 2D nitrides.
      Due to the quantum confinement, the reduced screening and, hence, the huge exciton binding energies, the sheet crystals appear as promising materials for active optoelectronic applications.

\begin{figure}[ht]
	      \includegraphics[scale=0.9]{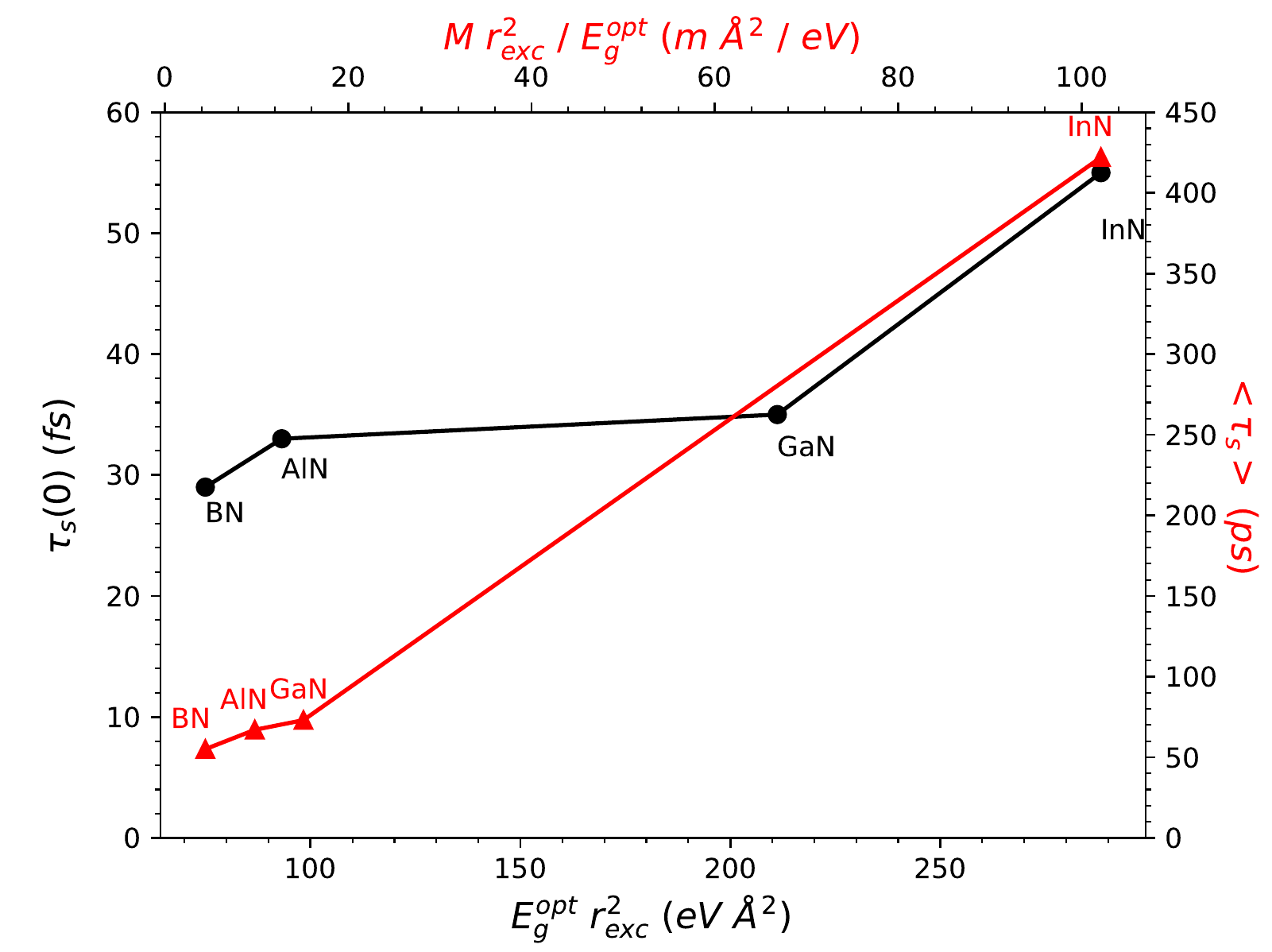}
	      \caption{
		      Excitonic lifetimes $\tau_S(0)$ and $\langle\tau_S\rangle$ at 300~K of the 2D nitrides versus characteristic exciton parameters.
		      }\label{binding_vs_lifetime}
      \end{figure}

      Finally, for a better understanding of the exciton radiative decay rates, we also apply the variational solutions for the lowest 1\emph{s} excitons with a pure 2D screening derived in Ref. \cite{pulci2015excitons}.
      With the parameters summarized in Table~\ref{tab:geo} we find the values given in parenthesis in columns 8 and 9..
      They are close to those computed \emph{ab-initio}.
      These findings indicate that the material dependence, the chemical trend, can be described by $\tau_S(0) \sim E_g^{opt}r^2_{exc}/|\bra{v {\bf k}}p_{\parallel}\ket{c {\bf k}}|^2$.
      For the thermally averaged quantities it holds $\langle \tau_S\rangle \sim \tau_S(0)M/(E_g^{opt})^2$.
      These relations are in qualitative agreement with the variation of the \emph{ab-initio} exciton lifetimes with the 2D crystal discussed above.
      However, Fig.~\ref{binding_vs_lifetime} also clearly indicates that the material dependence on the optical matrix element square $|\bra{v{\bf k}}p_\parallel \ket{c{\bf k}}|^2$ is not negligible, at least going from GaN to InN.
      Along the row BN, AlN and GaN, this dependence is weak.
      The exciton lifetime $\langle \tau_S\rangle$ at 300 K computed to about 73 (36) ps within the \emph{ab initio} (model) scheme for the first exciton in GaN (see also Table~\ref{tab:geo}) has to be related to similarly estimated values of 20 ps \cite{peng2018room} or 600 ps \cite{sanders2017electronic}.
      The weak difference in the first case \cite{peng2018room} is astonishing because of the used different approaches.
      The stronger deviation in the second case \cite{sanders2017electronic} is probably due to the applied larger exciton mass M estimated by bulk particle values and the different optical transition matrix elements taken into account.

  \section*{\label{sec4}Summary and Conclusions}
    In conclusion, the optoelectronic properties of group-III nitride monolayers have been here investigated \emph{ab-initio}.
    The quasiparticle electronic structures are computed within the $G_0 W_0$ approximation.
    The two-particle excitations and the accompanying optical properties are predicted by solving the Bethe-Salpeter equation with screened electron-hole attraction and electron-hole exchange.
    In addition to the absorption of the lowest bound electron-hole pairs, their emission properties are also studied by computing the exciton radiative lifetimes.
    The characteristic parameters of the bound excitons dominating both absorption and emission have been also investigated in a variational framework by applying static sheet polarizabilities to describe the screening and the effective mass approximation.
    \emph{Ab-initio} and model excitonic parameters are compared.
We have demonstrated a significant influence of the electron-hole interactions on the optical spectra of the 2D nitrides, in particular on the absorption spectra.
    Strong bound exciton peaks appear below the direct quasiparticle gap.
    The lowest energy excitons exhibits huge exciton binding energies of $0.6-2.0$~eV going from InN to BN.
    Thereby, the average electron-hole distances vary between $3.8$ and $15.5$~{\AA} from BN to InN, indicating a transition from a Frenkel-like to a Wannier-Mott-like behavior along the row BN, AlN, GaN, and InN.

    The emission properties of the 2D nitrides are illustrated by the radiative lifetimes of the lowest-energy excitons.
    We observed lifetimes of the order of $30-60$~fs, much faster than for other 2D semiconductors.
    In the case of the thermal average of the excitons with finite translation vector, the radiative lifetimes increase substantially.
    Their low values indicate that group-III nitride sheets appear to be promising materials for active optoelectronic applications in the visible up to the vacuum ultraviolet region.

  \section*{\label{sec2}Methods}
    % subsection{\label{sec2a}Ground and excited states}
      Our \emph{ab-initio} calculations of the structural and electronic properties proceed in three steps: first, we calculate the equilibrium geometry of 2D honeycomb sheets within density-functional theory (DFT) \cite{kohn1965self} as implemented in the \emph{Quantum Espresso} code \cite{giannozzi2009quantum,giannozzi2017advanced}, using the local density approximation (LDA) to describe exchange and correlation (XC).
      Minimizing the total energy with respect to the atomic coordinates the lattice constant and other geometry parameters are derived.
      The isolated 2D crystals are modeled within a supercell with a vacuum layer of thickness $L$ of about 15 \r{A} between the periodic images. Second, the single-quasiparticle excitation energies are derived by correcting the Kohn-Sham (KS) eigenvalues $\varepsilon^{KS}_{n{\bf k}}$
      of the DFT \cite{kohn1965self}.
      We  accurately determine the electronic levels by solving the quasiparticle (QP) equation within the G$_0$W$_0$ approximation for the XC self-energy \cite{bechstedt2016many} as implemented in the \emph{chisig} code \cite{chisig.x}.
      For the exchange part of the self-energy, we use a $102\times 102\times 1$ $\mathbf{k}$-point mesh centered on $\Gamma$ in the BZ.
      For the correlation part of the self-energy and for the screened Coulomb interaction W, we use a $51\times 51\times 1$ $\mathbf{k}$-point mesh and 300 bands.
    \subsection*{\label{sec2b}Electron-hole-pair states}
    In the third step, we account for the electron-hole interaction, calculating the screened electron-hole attraction and the unscreened electron-hole exchange.
      We construct the corresponding electron-hole pair Hamiltonian $H_{exc}$ and solve the homogeneous Bethe-Salpeter equation (BSE) \cite{bechstedt2016many}.
      Applying the \emph{dp4exc} code \cite{dp4exc} the eigenstates $|S{\bf Q}\rangle$ with set of quantum numbers $S$, translational momentum $\hslash{\bf Q}$, and energy $E_S({\bf Q})$ are determined within the Tamm-Dancoff approximation.

      For a better understanding of the lowest-energy ground state excitons with $S=0$, ${\bf Q}=0$ and excitation energy $E_0(0)$ with an approximate 1$s$ character, we compare the \emph{ab-initio} results for the exciton binding energy $E_b$ and the exciton radius $r_{\rm exc}$ obtained within the accurate, but computationally heavy BSE calculations, with predictions of an analytical model of excitons in two dimensions \cite{cudazzo2011dielectric, pulci2015excitons}.
      We focus on the bound excitons at the absorption edge.
      Their Schr\"odinger equation is formulated  within the effective mass approximation (EMA), where the reduced exciton mass $\mu=m_em_h/(m_e+m_h)$ is introduced with the electron (hole) mass $m_e$ ($m_h$) derived from dispersion of the lowest conduction (highest valence) band.
      The screened Coulomb interaction between electron and hole is described by a Keldysh potential \cite{Keldysh:1979:JETPL,pulci2015excitons} whose distance dependence is ruled by 
      the 2D static electronic polarizability of the sheet,  $\alpha_{2D} = L \cdot ({\text Re} \epsilon _{\parallel}(\omega = 0) - 1) / 4 \pi$, which is computed from the in-plane dielectric function $\epsilon _{\parallel}$ in the limit of vanishing wave vector and frequency.
      The lowest exciton state is determined by a variational approach.

    \subsection*{\label{sec2c}Optical properties}
      With the solutions $|S{\bf Q}\rangle$ and $E_S({\bf Q})$ of the BSE, the frequency-dependent dielectric tensor $\epsilon_{\|/\bot}(\omega)$ can be directly computed for light polarization parallel/perpendicular to the crystal sheets in the superlattice arrangement.
      Because of the neglect of photon wave vector, only vertical optical transitions appear with ${\bf Q}=0$ and optical transition matrix elements $\langle c{\bf k}|p_{\|/\bot}|v{\bf k}\rangle$ of the momentum operator between KS valence $(|v{\bf k}\rangle)$ and conduction $(|c{\bf k}\rangle)$ states.
      The optical properties of isolated sheet crystals are described by the optical conductivities $\sigma_{\|/\bot}(\omega)$ \cite{matthes2016influence}.
      Their real (imaginary) parts describe the absorption (dielectric) properties for in- and out-of-plane light polarization.

      Following \cite{moody2016exciton}, we calculate the radiative decay rate $\gamma_S(0)$ and lifetime $\tau_S(0)$ of an exciton in state $S$ with vanishing wavevector ${\bf Q}=0$ as \cite{palummo2015exciton}
      \begin{equation}\label{eq:tau$_s$(0)}
      	\gamma_S(0)    = 
        \tau^{-1}_S(0) =
        \frac
          {
            8\pi e^2E_S(0)
          }
          {
            \hslash^2c
          }
          \frac
          {
            \mu_S^2
          }
          {
            A_{uc}
          },
      \end{equation}
      where $\mu_S^2$ is the square modulus of the exciton dipole transition matrix element.
      $A_{uc}$ is the area of the unit cell and $E_S(0)$ is the exciton excitation energy.
      To compute the thermal average radiative lifetimes $\langle\tau_S\rangle$ of excitons in excitonic bands $E_S$({\bf Q}) at temperature $T$ we apply the formula obtained in the classical high-temperature limit \cite{palummo2015exciton}
      \begin{equation}\label{eq:tau$_s$}
      	\langle 
          \tau_S
        \rangle = 
        \tau_S(0)
        \frac{3}{4}
        \left(
          \frac{E_S^2(0)}{2Mc^2}
        \right)^{-1} k_BT
      \end{equation}
      with $M=m_e+m_h$ being the translational (total) mass of the exciton.
  \bibliography{bibliography}
  \section*{Acknowledgments}
    % section{Acknowledgments}
    O.P. and I.K. acknowledge financial support from the EU MSCA HORIZON2020 projects 'CoExAN' (GA644076) and DiSeTCom (GA 823728). F.B. acknowledges travel support by INFN Tor Vergata.
   We thank Prof. Maurizia Palummo for useful discussions.

  \section*{Author contributions statement}
    M.S.P, D.G. and O.P. performed the calculations.
    F.B, V.O.  and O.P. conceived the project.
    All authors wrote the paper.
    All authors analyzed the results.
    All authors reviewed the manuscript.

  \section*{Additional information}
    The Authors declare no competing financial interest.
    
       \begin{table*}[!h]
	      \begin{tabular}{|c|c|c|c|c|c|c|c|c|}
	      \hline \hline
	          & $a$ ({\AA}) & $E_g$ (eV)                            & $E_b$ (eV) & $r_{exc}$ ({\AA}) & $\alpha_{2D}$ (a.u.) & $m_h/m_e$ ($m$) & $\tau_S(0)$ (fs) &$ \langle \tau_S \rangle$(ps) \\ \hline
	      BN  & 2.48        & 7.2 (KK)/6.7 (K$\Gamma$)              & 2.0 (2.1)  & 3.8  (3.8)        & 2.10                 &  0.63/0.96   & 29 (13)         & 55 (14)    \\ \hline
	      AlN & 3.03        & 6.5 ($\Gamma\Gamma$)/5.8 (K$\Gamma$)  & 1.9 (2.2)  & 4.5  (3.4)        & 2.07                 &  1.64/0.59   & 33 (1)          & 67 (18)        \\ \hline
	      GaN & 3.15        & 4.5 ($\Gamma\Gamma$)/ 4.6 (K$\Gamma$) & 1.2 (1.4)  & 8.0  (6.6)        & 2.78                 &  0.52/0.26    &  35 (18)         & 73 (36)   \\ \hline
	      InN & 3.52        & 1.7($\Gamma\Gamma$)/2.1 (K$\Gamma$)   & 0.6 (0.5)  & 15.5 (16.4)       & 7.76                 &  0.42/0.09  &  55 (58)         & 422 (445)    \\ \hline \hline
	      \end{tabular}
	      \caption{
		      Lattice constant $a$ and direct and indirect QP gaps $E_g$.
		      The band edge positions in the BZ are indicated in parenthesis.
          Excitonic binding energy $E_b$, excitonic radius $r_{exc}$.
          Also the DFT ingredients $\alpha _{2D}$ and the electron and hole effective masses are reported.  Last two columns: calculated exciton radiative lifetimes for 2D III-nitrides $\tau_S(0)$ (\ref{eq:tau$_s$(0)}) and $\langle \tau_S\rangle$ (\ref{eq:tau$_s$}) for room temperature and bound excitons in the ground state $S=0$ and ${\bf Q}$=0.
          The values in parenthesis have been estimated within the effective mass approximation but \emph{ab-initio} calculated optical matrix elements.
		      }\label{tab:geo}
      \end{table*}

    
\end{document}

  %\preprint{APS/123-QED}

  \title{Supplemetal Material for "Giant excitonic absorption and emission in two-dimensional group-III nitrides"}% Force line breaks with \\
  \author[1]{Maria Stella Prete}
  \author[1]{Davide Grassano}
  \author[1,*]{Olivia Pulci}
  \author[2]{Ihor Kupchak}
  \author[3]{Valerio Olevano}
  \author[4]{Friedhelm Bechstedt}
  \affil[1]{ Dipartimento di Fisica, Universit\`{a} di Roma Tor Vergata, and INFN, Via della Ricerca Scientifica 1, I-00133 Rome, Italy}
  \affil[2]{
  V.E. Lashkaryov Institute of Semiconductor Physics, National Academy of Sciences of Ukraine, Kyiv, Ukraine}
  \affil[3]{CNRS, LPMMC, 38042 Grenoble, France}
  \affil[4]{ IFTO, Friedrich Schiller Universit\"at, Max-Wien Platz 1, D-07743 Jena, Germany}
  \affil[*]{olivia.pulci@roma2.infn.it}

  \begin{abstract}
    Absorption and emission of pristine-like semiconducting monolayers of BN, AlN, GaN, and InN are here systematically studied by \emph{ab-initio} methods.
    We calculate the absorption spectra for in-plane and out-of-plane light polarization including quasiparticle and excitonic effects.
    Chemical trends with the cation of the absorption edge and the exciton binding are  discussed in terms of the band structures.
    Exciton binding energies and localization radii are explained within the Keldysh model for excitons in two dimensions.
    The strong excitonic effects are due to the interplay of low dimensionality, confinement effects, and reduced screening.
    We find exciton radiative lifetimes ranging from tenths of picoseconds (BN) to tenths of nanoseconds (InN) at room temperature, thus making 2D nitrides, especially InN,  promising materials for light-emitting diodes and high-performance solar cells.
  \end{abstract}

  \date{\today}% It is always \today, today,
  \maketitle

  %\setcounter{equation}{0}
  %\setcounter{figure}{0}
  %\setcounter{table}{0}
  %\setcounter{page}{1}
  %\makeatletter
  \renewcommand{\theequation}{SM\arabic{equation}}
  \renewcommand{\thefigure}{SM\arabic{figure}}

  \section*{\label{sec1}Quasiparticle Electronic Structure}

    \begin{figure*}[ht]
      \includegraphics[scale=0.70]{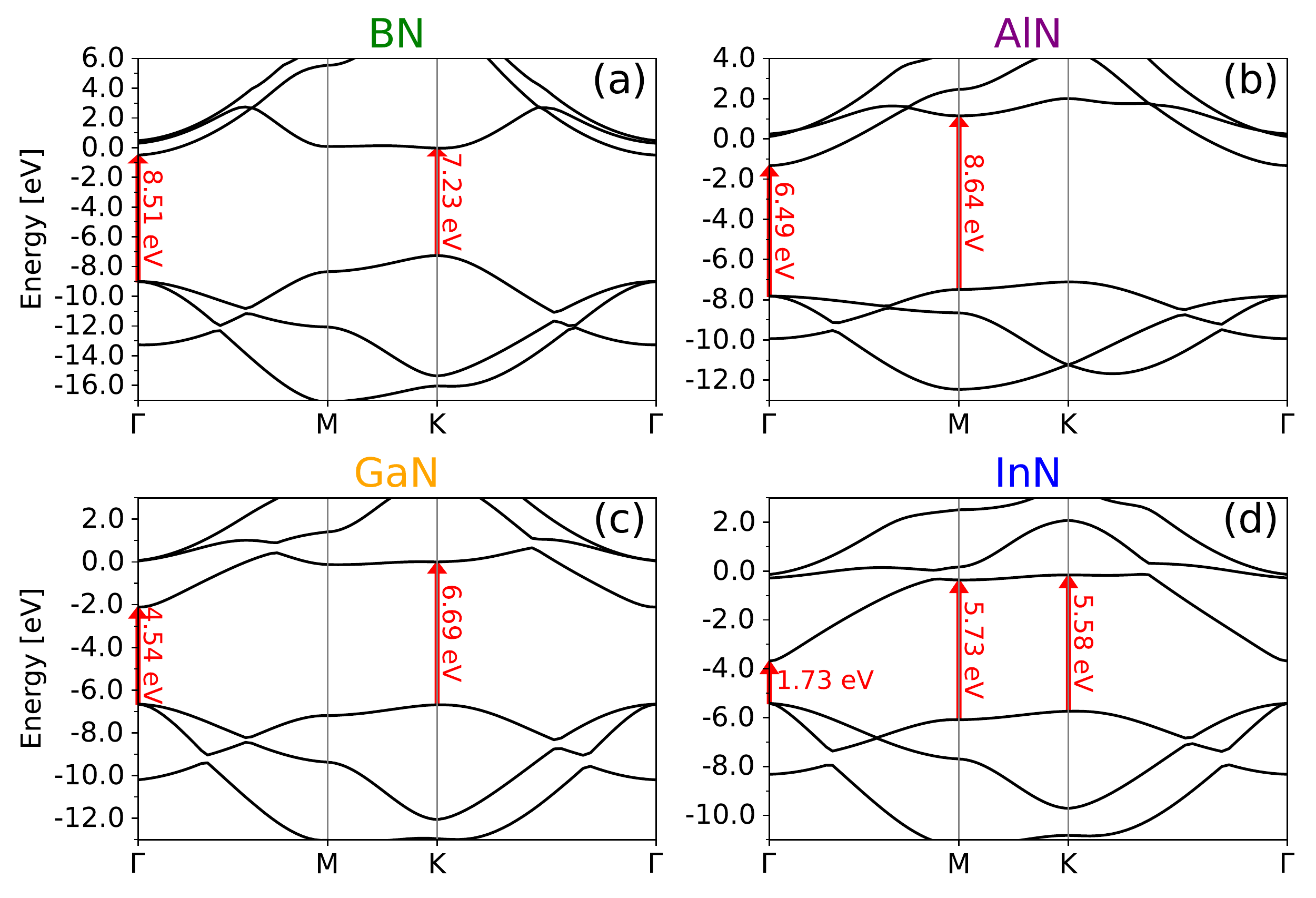}
      \caption{
        Electronic quasiparticle band structures of (a) BN, (b) AlN, (c) GaN and (d) InN sheet crystals obtained within the one-shot G$_0$W$_0$ approximation.
        The vertical arrows indicate possible strong optical transitions in these 2D materials.
        The vacuum level position is used as energy zero.
        }\label{fig:SM1}
    \end{figure*}

    The electronic band structures of the 2D group-III nitrides are displayed in Fig.~\ref{fig:SM1}.
    Although their geometry resumbles that of the graphene, the different electronegativities between nitrogen and B, Al, Ga and In atoms open significant gaps at the $K$ points of the Brillouin zone (BZ).
    The conduction band minimum moves to the $\Gamma$ point for all materials studied.
    In the case of the light cation, i.e., BN and AlN, the valence band maximum is fixed at $K$ (as in graphene).
    However, in the case of the heavier cations, Ga and In, it also moves to the $\Gamma$ point.

    Direct allowed optical transitions are indicated in Fig.~\ref{fig:SM1} by vertical arrows for not too large interband energies.
    The lowest one occurs at $\Gamma$, independent of the cation.
    Higher-energy transitions are however possible at $M$ and $K$ between the uppermost valence band and the lowest conduction band.
    Transitions into the second-lowest conduction band may also occur at $\Gamma$, in particular for BN and AlN, where the energy distance to the conduction band minimum is relatively small compared to the gap size.

  \section*{\label{sec2}Optical Spectra}

    \begin{figure*}[ht]
      \includegraphics[scale=0.38]{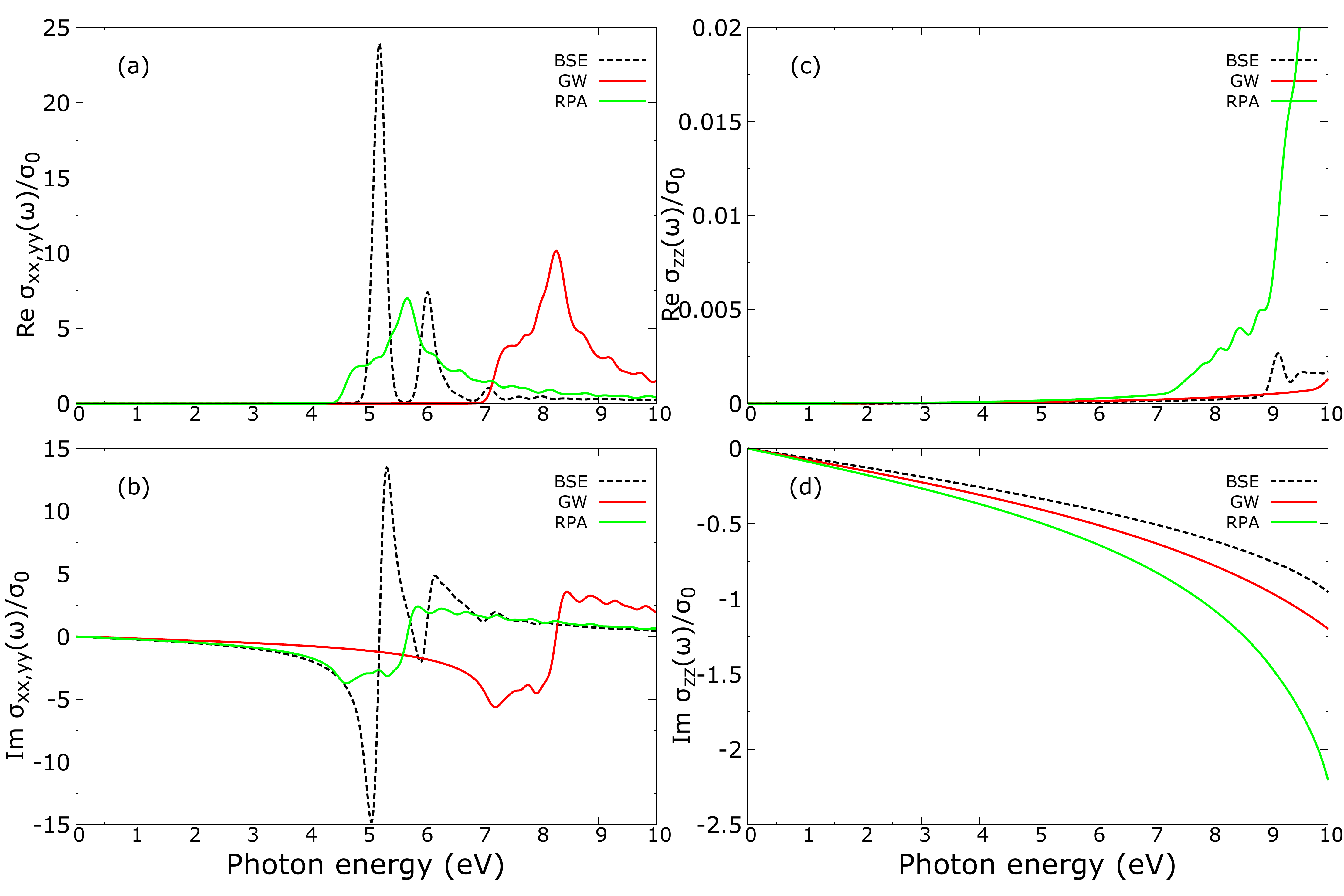}
      \caption{
        Real (a,c) and imaginary (b,d) parts of optical conductivity for BN calculated within the independent-particle approach (RPA - green lines), within the independent-quasiparticle approach (GW - red lines) and taking into account excitonic effects (BSE - black dotted lines) for in-plane (a,b) and out-of-plane polarization (c,d).
        }\label{fig:BN}
    \end{figure*}
    \begin{figure*}[ht]
      \includegraphics[scale=0.38]{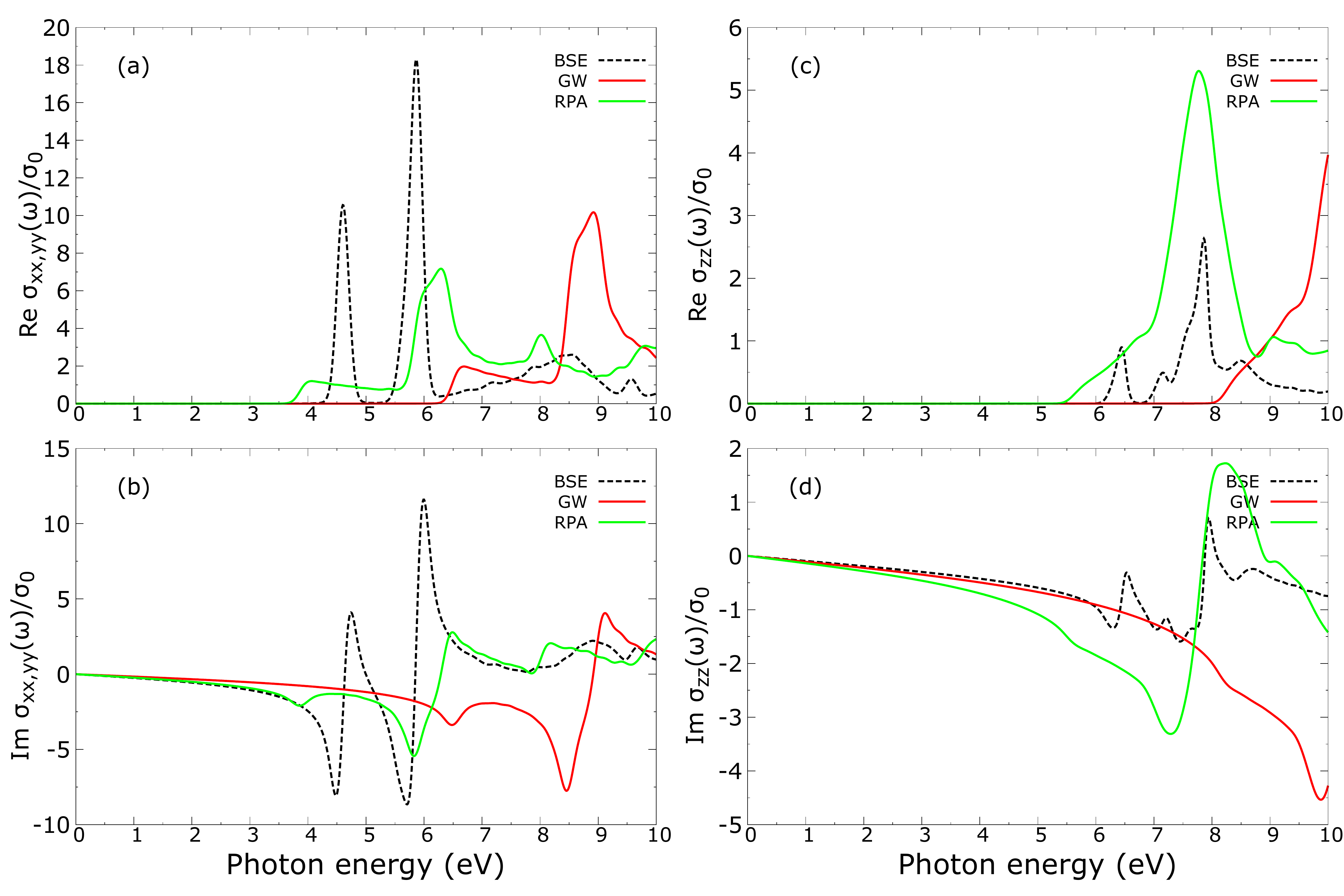}
      \caption{
        Real (a,c) and imaginary (b,d) of optical conductivity for AlN calculated within the independent-particle approach (RPA - green lines), within the independent quasiparticle approach (GW - red lines) and taking into account excitonic effects (BSE - black dotted lines) for in-plane (a,b) and out-of-plane polarization (c,d).
        }\label{fig:AlN}
    \end{figure*}
    \begin{figure*}[ht]
      \includegraphics[scale=0.38]{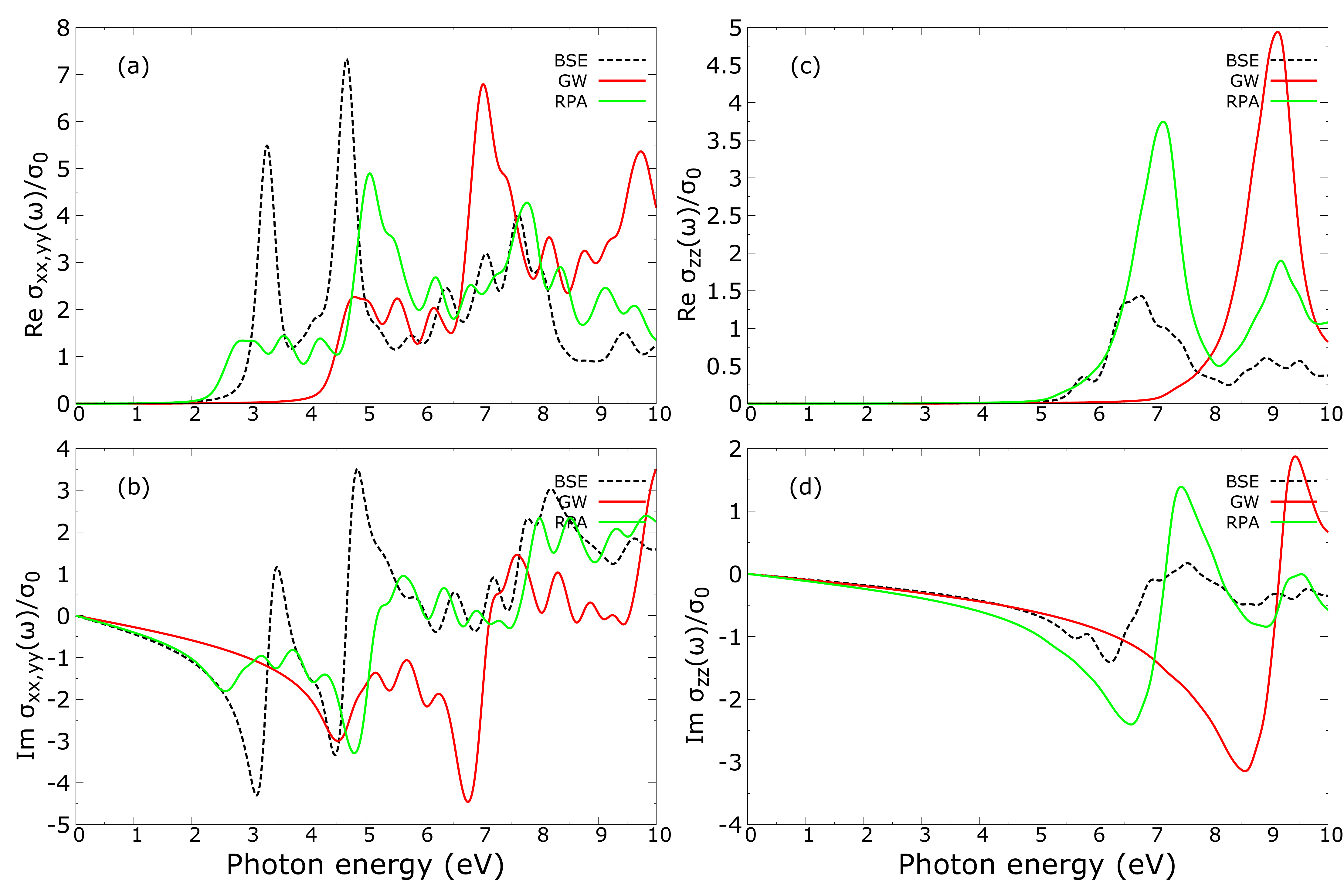}
      \caption{
        Real (a,c) and imaginary (b,d) of optical conductivity for GaN calculated within the independent-particle approach (RPA - green lines), within the independent quasiparticle-approach (GW - red lines) and taking into account excitonic effects (BSE - black dotted lines) for in-plane (a,b) and out-of-plane polarization (c,d).
        }\label{fig:GaN}
    \end{figure*}
    \begin{figure*}[ht]
      \includegraphics[scale=0.38]{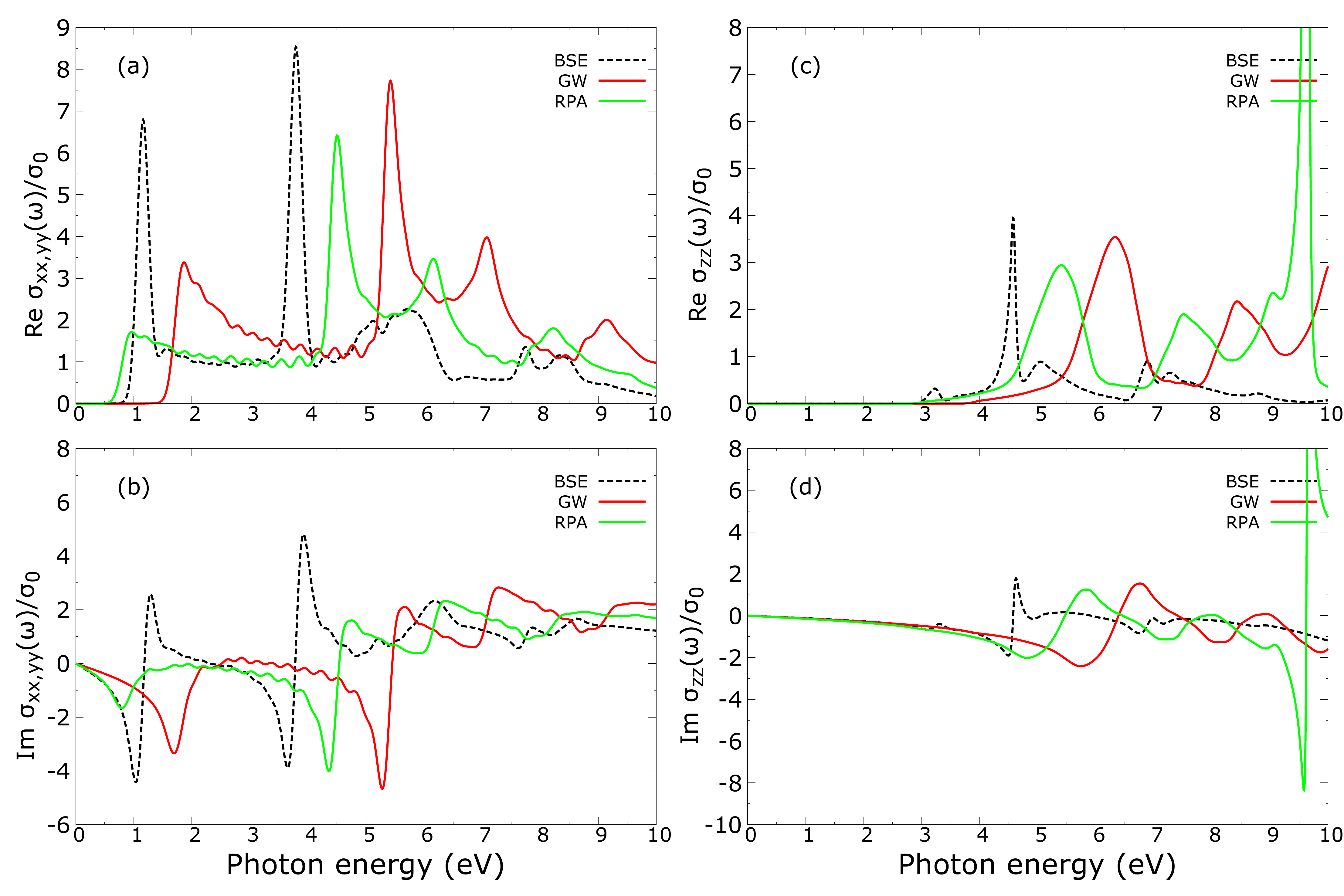}
      \caption{
      Real (a,c) and imaginary (b,d) of optical conductivity for InN calculated within the independent-particle approach (RPA - green lines), within the independent quasiparticle-approach (GW - red lines) and taking into account excitonic effects (BSE - black dotted lines) for in-plane (a,b) and out-of-plane polarization (c,d).
      }\label{fig:InN}
    \end{figure*}

    In order to characterize the linear optical properties of the monolayer nitrides, the real and imaginary parts of the 2D optical conductivity are presented in Figs.~\ref{fig:BN} -- \ref{fig:InN} for both in-plane and out-of-plane light polarization.
    The many-body effects are described in three different approximations:
    \begin{enumerate}
      \item[(i)]
        In the independent-particle approximation the electron-hole interactions are not taken into account.
        The single-particle eigenfunctions and eigenenergies are taken from the Kohn-Sham (KS) approach.
      \item[(ii)]
        In the independent-quasiparticle (QP) approach the KS eigenvalues are corrected by QP shifts computed within the G$_0$W$_0$ approximation for exchange and correlation interactions.
        Correspondingly, the spectra are significantly blueshifted by about 2.5~eV (BN, AlN), 1.8~eV (GaN), and 1.0~eV (InN).
      \item[(iii)] 
        The BSE spectra including the screened electron-hole attraction and the unscreened electron-hole exchange interaction are characterized by three important features with respect to the independent-QP approach.
        There is, in general, a redshift accompanied by a spectral redistribution.
        The most striking feature is, however, the formation of bound exciton states below the QP gaps.
    \end{enumerate}
    These phenomena are visible in the real parts or the optical conductivity, especially by significant peaks, while in its imaginary parts still some resonance features are observable.
    The real parts vanish for small photon energies.
    Their slopes characterize the dielectric properties of the sheets.

    The most important feature of changing the light polarization from an in-plane position to the (normal) out-of-plane direction is a significant blueshift of the spectra.
    It can be related to depolarization effects or local-field effects as described by the electron-hole exchange interaction in the pair Hamiltonian $H_{\rm exc}$.